\documentclass[prl,twocolumn,aps,showpacs,floatfix]{revtex4}

\usepackage{graphicx}
\usepackage{dcolumn}
\usepackage{bm}
\begin{document}
\preprint{}

\title{Intrinsic and extrinsic mirror symmetry breaking in anti-dot spin-wave waveguides}

\author{J. W. K{\l}os$^1$}\thanks{These authors contributed equally to this work.}
\author{D. Kumar$^2$}\thanks{These authors contributed equally to this work.}
\author{M. Krawczyk$^1$}
\email{krawczyk@amu.edu.pl}
\author{A. Barman$^2$}
\email{abarman@bose.res.in}
\affiliation{$^1$Faculty of Physics, Adam Mickiewicz University in Poznan, Umultowska 85, Pozna\'{n}, 61-614, Poland;\\ $^2$Department of Condensed Matter Physics and Material Sciences, S. N. Bose National Centre for Basic Sciences,\\ Block JD, Sector III, Salt Lake, Kolkata 700 098, India}

\date{\today}

\begin{abstract}
We theoretically study the spin-wave spectra in magnonic waveguides periodically patterned with square anti-dots in nanoscale with pinned magnetization at the edges.  We show that the breaking of the mirror symmetry of the waveguide by the structural changes can result in a magnonic band gap closing.  These intrinsic symmetry breaking can be compensated by properly chosen asymmetric external bias magnetic field, i.e. in  an extrinsic way. As a result the magnonic gaps existing in the ideal symmetric structure can be recovered. The model used for the explanation also suggests that this idea could be generic both for exchange and dipolar interaction regimes of spin-waves and also for other types of waves, e.g., electrons in the graphene ribbons.

\end{abstract}
\pacs{75.30.Ds,75.75.-c,75.78.Cd,75.70.Ak}
\maketitle


The possibilities for fabrication metallic magnetic materials with nanoscale precision has opened the way for tailoring the dispersion of high-frequency spin waves (SWs). This can be done also with the use of  magnonic crystals (MCs), which are  magnetic counterparts of photonic crystals.\cite{Puszkarski03,Nikitov01} Recently the first  bi-component MCs have been realized at the nanoscale  and the opening of magnonic gaps were proved experimentally.\cite{Wang10, Tacchi12}  Two-dimensional (2D) anti-dots lattices (ADL) formed by periodic array of holes in ferromagnetic film can be much easily fabricated. These systems have been intensively studied in recent years on  length scales from micrometers down to few tens of nanometers.\cite{Pec05,Neus08,Neus10,Tacchi10,Hu11,Mandal12} For ADLs with big period the inhomogeneity of the internal magnetic field is decisive for formation of the magnonic band structure.\cite{Zivieri12} With the decreasing period of ADL the Brillouin zone (BZ) border will move to larger wave-vectors and the exchange interactions at some point start to play a primary role in the formation of a magnonic band structure.

Prototypes of basic magnonic devices have already been demonstrated to be promising for technological applications\cite{Kru10b,Serga10,Khitun12}  but the scaling down of magnonic elements to tens of nanometers in size and tens of GHz operating frequencies\cite{Bonetti09, Prokopenko11, Madami11} are still a challenge.    Waveguides for SWs, which are  an important component in most magnonic devices,\cite{Kru10b} have been realized experimentally so far only for  SWs in the frequency range up to few GHz.\cite{Kozhanov09,Neus10, Demidov11,Clausen11,Duer12,Mamica12} To predict properties of magnonic devices at scale of tens of nm, more basic research needs to be conducted. Therefore, theoretical investigation of the SW waveguides operating in the range of tens of GHz is a frontier field of research. 

Recently the micromagnetic simulations (MSs) were used to show that periodic waveguides have filter properties due to the opening of magnonic gaps in the SW spectrum.\cite{Lee09, Ma11,Kim10,Bai11} 
In this Letter we investigate how loss of a mirror symmetry  within a one-dimensional (1D) nanoscale anti-dot lattice waveguide (ADLW) may affect the SW dispersion. When this symmetry exists, based on their profile with respect to the central longitudinal axis, the SW spectra can be separated into two groups: symmetric modes and anti-symmetric modes. The breaking of the mirror symmetry will automatically make the classification impossible. We will study two types of the symmetry breaking mechanisms namely, intrinsic and extrinsic, which are realized by changes of the position of the anti-dots row and by application of asymmetric bias magnetic field, respectively.  The question is how do these changes influence the magnonic spectra and the existing magnonic gaps? How big the symmetry breaking needs to be in order to close the gap?  We  perform theoretical calculations with the plane wave method (PWM) and confirm the results by using MSs, to address these questions. The discussion on the possible extensions of our findings to large scale and other systems will be conducted on basis of an analytical model.



\begin{figure}[!ht]
\includegraphics[width=7 cm]{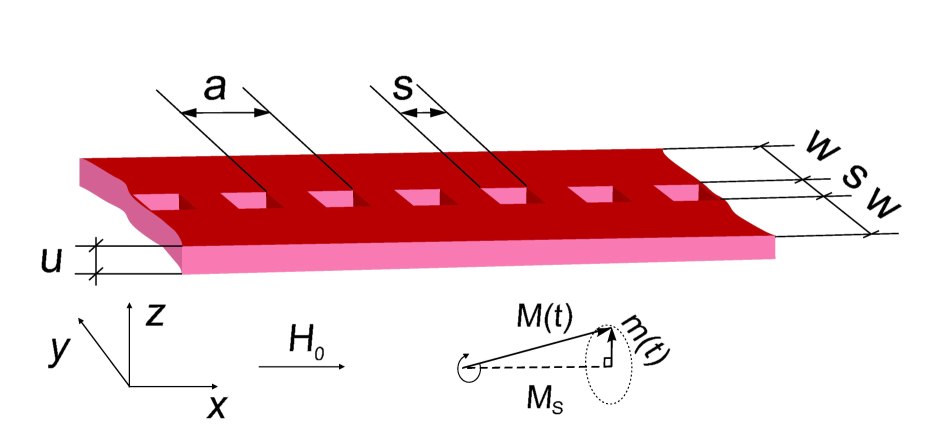}
\caption{Anti-dots lattice waveguide under investigation: $u =$ 3 nm thick and $2w+s =$ 45 nm wide infinite Py stripe with a periodic series of square anti-dots (of size $s\times s$, where $s$ = 6 nm) disposed along the waveguide with a period of $a =$ 15 nm. The row of anti-dots divides  the waveguide into two sub-waveguides of the width $w$  = 19.5 nm each. Bias magnetic field $\mu_0H_0$ = 1 T is oriented along the waveguide.}
\label{Fig:waveguide}
\end{figure}

We  study the symmetric and asymmetric magnonic waveguides based on the anti-dots lattice structure shown in Fig.~\ref{Fig:waveguide}. It has a form of a thin (thickness $u =$ 3 nm) and infinitely long permalloy (Ni$_{80}$Fe$_{20}$)  stripe with a single row of square holes ($s = 6$ nm anti-dots) disposed periodically along the central line. The stripe  width and the lattice constant are fixed at $2 \times w + s= 45$ nm and $a=$ 15 nm, respectively.  The row of holes is placed in the distance $w =$ 19.5 nm from the top (and bottom) edge of the stripe in the case of the symmetric ADLW. A bias magnetic field is applied along the stripe and it is strong enough  to saturate the sample ($\mu_{0} H_{0}=1$ T). 
The saturation magnetization $M_{\text{s}} = 0.8 \times 10^{6}$ A/m, the exchange length $\lambda_{\text{ex}} = 5.69 $ nm, and the gyromagnetic ratio $\gamma$ = 175.9 GHz/T were assumed in calculations. 

The calculations of the magnonic band structure are performed with the finite difference method MS and the PWM, with OOMMF\cite{Donahue99}  and a  Fortran code developed by authors, respectively. Both methods solve the Landau-Lifshitz-Gilbert (LLG) equation:
\begin{eqnarray}
\frac{\partial {\bf M}({\bf r},t)}{\partial t}&=&\gamma\mu_{0}{\bf M}({\bf r},t)\times {\bf H}_{\text{eff}}({\bf r},t)\\ \nonumber && -\frac{\alpha}{M_{\text{s}}}\left({\bf M}({\bf r},t)\times\frac{\partial {\bf M}({\bf r},t)}{\partial t}\right),\label{eq:ll}
\end{eqnarray}
where ${\bf r}$ and $t$ are position vector and time, respectively. $\mu_{0}$ is the permeability of vacuum. The first term on the right hand side is related to the torque inducing precession of the magnetization ${\bf M}$ and the second one describe the damping process ($\alpha$ denotes the damping constant). The damping is neglected in PWM calculations and included in MS ($\alpha=$ 0.0001). The effective magnetic field ${\bf H}_{\text{eff}}$ here consists of the bias magnetic field $H_{0}$, demagnetizing field and exchange field. The pinned dynamical components of the magnetization vector were assumed at Py/air interfaces in calculations with both methods. The pinning in OOMMF was introduced by fixing magnetization vector in all cells of the discretization mesh which border the anti-dots, i.e., for the width 0.5 nm along $y$ axis.
(In MS the discrete mesh size of $1.5 \times 0.5 \times 3$ nm along $x$, $y$ and $z$ axis, respectively, were used. The MS were performed for 4 ns. In the PWM we use 961 plane waves.)
In the PWM the pinning is applied exactly at the edges of Py. Due to small thickness of the ADLW, uniform SW profile across the thickness is assumed.  
Both methods were already used in the calculations of the SW dynamics and proved to give correct results.\cite{Neus08,Tacchi10,Kumar12}


\begin{figure}[!ht]
\includegraphics[width=8.5 cm]{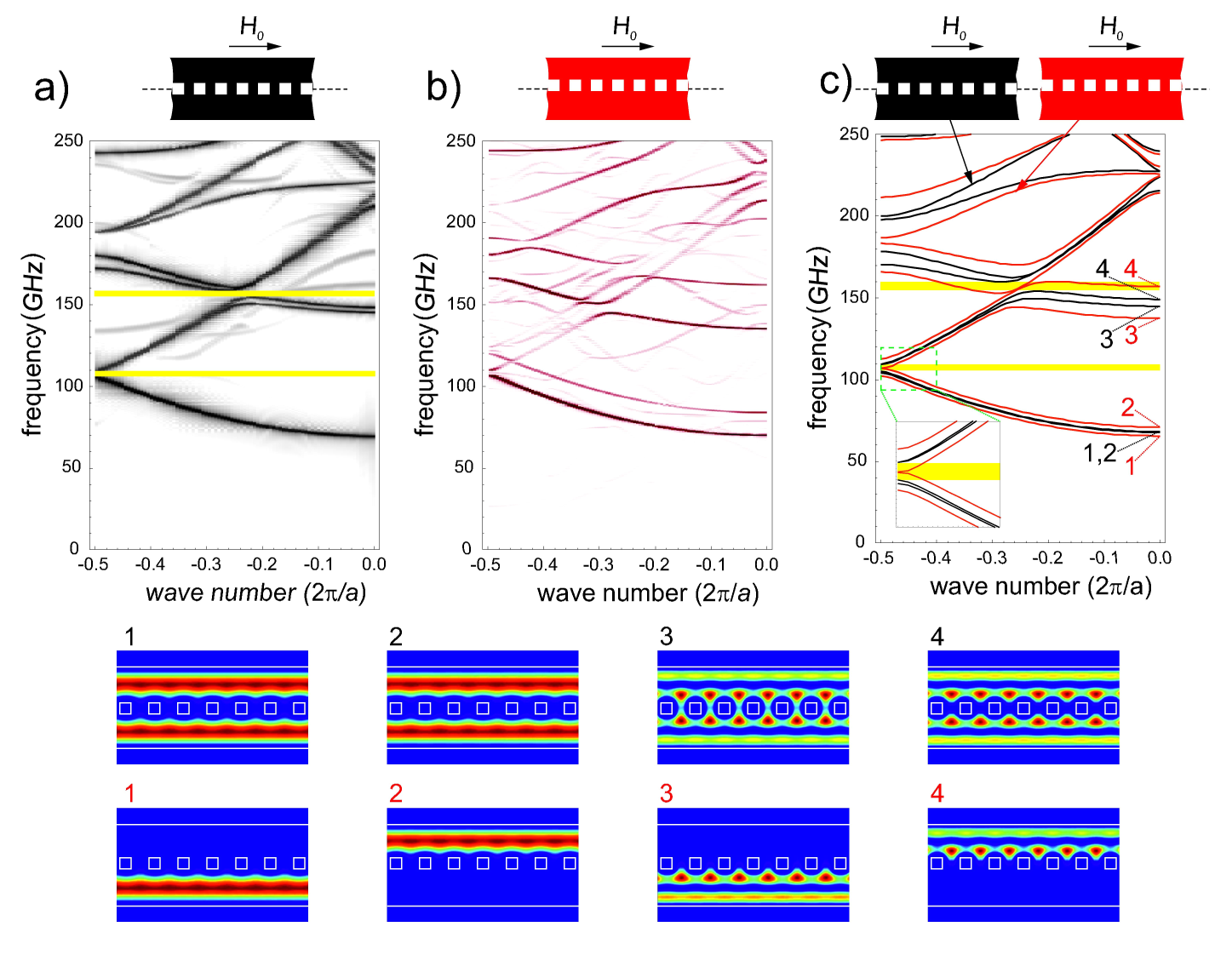}
\caption{(Color online) Magnonic band structure of ADLWs (shown in insets above the main figures where the thin dashed line marks the middle of the ADLW) calculated with OOMMF in (a) and (b), and with PWM in (c). The band structure for the symmetric ADLW are shown in (a) and in (c) with  black solid lines. The results for  ADLW with shifted  upward anti-dots row  are shown in (b) and in (c) with red lines calculated with OOMMF ($\Delta w$ = 1 nm) and PWM ($\Delta w$ = 0.9 nm), respectively. At the bottom the squared amplitudes of the dynamical magnetization $|m_z|^2$ for first four modes in the center of the BZ (calculated with PWM--cf. (c))  are shown for symmetric and asymmetric ADLW, first and the second row, respectively.}
\label{Fig:as_adw}
\end{figure}

 We start our investigation for the symmetric ADLW  (Fig.~\ref{Fig:waveguide}).\cite{Klos12} The dispersion relation of SWs in the symmetric ADLW is shown in Fig.~\ref{Fig:as_adw}, the results of the OOMMF simulations  are shown in Fig.~\ref{Fig:as_adw}(a) and of the PWM in Fig.~\ref{Fig:as_adw}(c) (black solid lines).  
The agreement between results of these two methods is satisfactory. The presence of two magnonic band gaps is evident and they are marked by the yellow color. The origin of these two band gaps was found to be different. The first one opens at the BZ boundary due to a Bragg reflection of SWs, while the second gap is opened in the middle  of the  BZ.\cite{Klos12} It was shown that this splitting of the bands is due to the  anti-crossing between two family of modes,\cite{Lee09} those with and without a nodal line in the upper and lower part of the ADLW (see the first row of profiles in Fig.~\ref{Fig:as_adw}). 


The structure investigated above has a mirror symmetry with respect to the central axis of the ADLW. The 1$^{\text{st}}$  and the 2$^{\text{nd}}$ mode (and also 3$^{\text{rd}}$ and 4$^{\text{th}}$) is symmetric and antisymmetric with respect to the central axis of the ADLW, respectively (see the first row of profiles at the bottom in Fig.~\ref{Fig:as_adw}) and their frequencies are degenerate. The shift of the row of anti-dots from the central line will break this symmetry. Frequencies of modes will split, with one mode shifted up and the second mode shifted down in the frequency scale. The dispersion relations of SWs in asymmetric ADLWs, obtained by shifting the row of anti-dots by $\Delta w =$ 1 nm and 0.9 nm upward and calculated using OOMMF and PWM, are respectively presented in Fig.~\ref{Fig:as_adw}(b) and in (c) (red lines).
(In OOMMF slightly larger value of $\Delta w$ were used because of the limitations of the discretization mesh and time needed for simulations.)
We see that the shift of the anti-dots row by only  2\% of the total width of the ADLW is enough to close both magnonic gaps.  The first mode, with lower frequency, is now concentrated in the wider part of the ADLW (width $w + \Delta w$) (see, Fig.~\ref{Fig:as_adw} in the bottom). The second mode, with higher frequency, has an amplitude concentreted in the upper part of the ADLW (of width $w - \Delta w$). The same mechanism is responsible also for closing the second gap, even though the origin of this gap is different and the respective changes of the frequencies of the third and the fourth bands are larger.
\begin{figure}[!ht]
\includegraphics[width=8.5 cm]{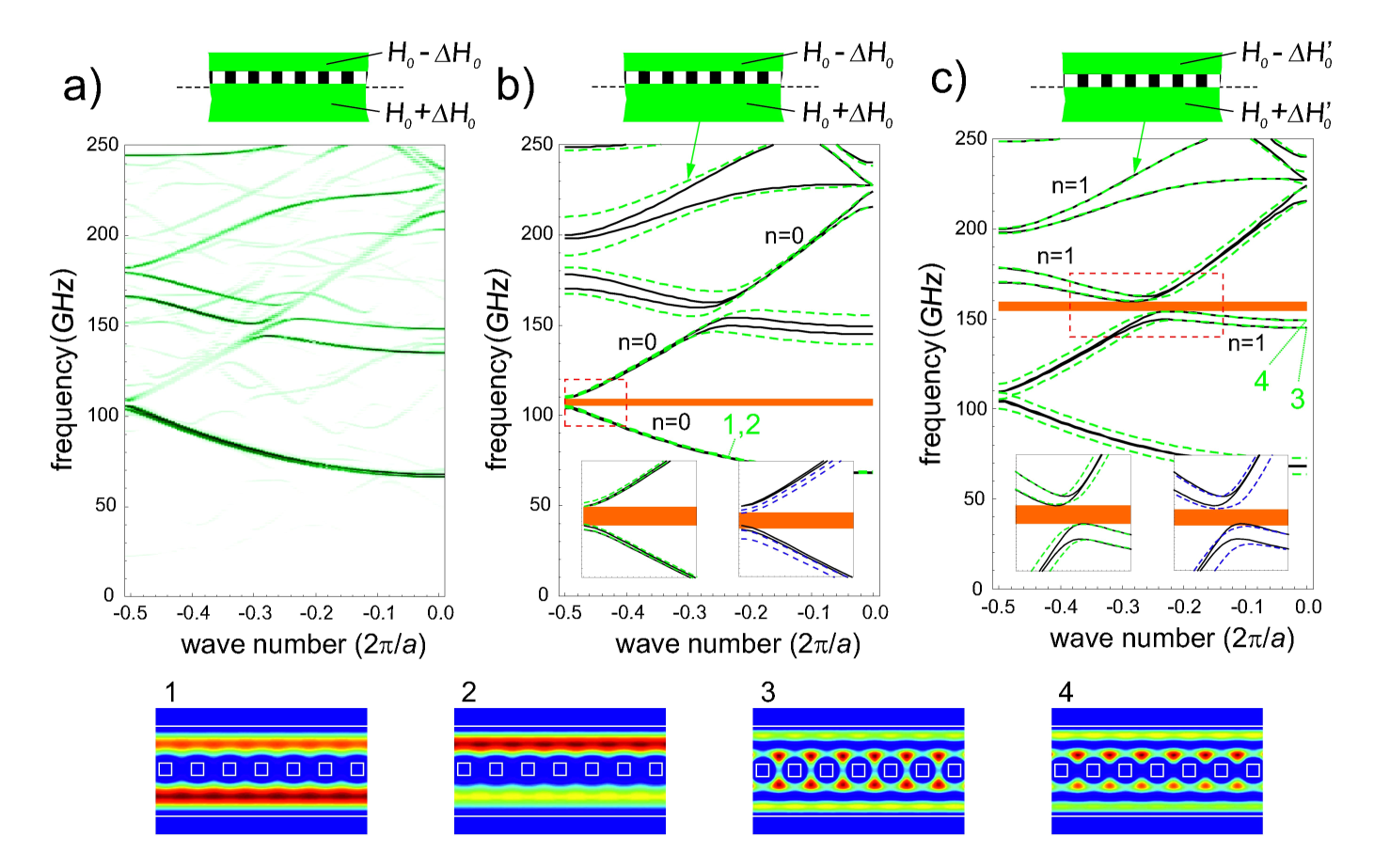}
\caption{(Color online) Width of magnonic gaps in ADLW as a function of (a) the shift of the anti-dots row for $\Delta w$ and (b) an additional asymmetric bias magnetic field $\Delta H_{0}$ in the symmetric ADLW. The $\Delta H_0$ increases bias magnetic field in the upper half of ADLW (to the value $H_{0} + \Delta H_{0}$) and decreases bias magnetic field below the row of anti-dots (to the value $H_{0} - \Delta H_{0}$).}
\label{Fig:Field}
\end{figure}

The gap widths as a function of the value of $\Delta w$ are shown in Fig.~\ref{Fig:Field}(a). The widths decrease monotonously with increasing $\Delta w$, the slope for the second gap is bigger and this gap is closed at 0.45 nm, while the first gap exists up to 0.8 nm.

We have just shown that the magnonic spectra and especially the magnonic gaps in the considered ADLW  are very sensitive to a mirror symmetry breaking. 
The question comes: is it possible to compensate the effect of an intrinsic symmetry breaking in the ADLW by the extrinsic one? In our case it will be a compensation of the effect of the structural asymmetry on the magnonic band structure (and magnonic gaps) by asymmetric bias magnetic field. The answer will begin from the consideration of the analytical model.



In the homogeneous waveguide (WG) the solutions of the linearized LLG equation (with damping neglected) can be written in the following form:
$
{\bf m}(x,y) = {\bf m}(y)  \text{e}^{i k_{x} x},
$
where $k_{x}$ is the wave-vector of the SW along the WG and $ {\bf m}(y)$ describe the dependence of the amplitude of dynamical components of the magnetization {\bf m} across the WG width (we assume the uniform magnetization across the WG thickness which is much less than the width). 
The solutions  can be estimated as:
$ {\bf m}(y) \approx \text{sin}(\kappa  y ),\;\text{cos}(\kappa  y )$
where the transversal component of the wave-vector $\kappa = (n+1) \pi/w_{\text{eff}}$ is quantized  ($n = $ 0, 1, 2, ... counts the number of nodal lines across the WG width). For strong but not ideal pinning the effective width $w_{\text{eff}} = w d/(d-2)$ depends on the pinning parameter $d$, which determines the boundary conditions (BC) for magnetization and gives also a possibility to include the dipolar effects into the model.\cite{Guslienko05} It  varies in general from $0$ to $\infty$ for the transition from unpinned to pinned BC. The pinning parameter
 $d=2\pi (1-\frac{K_{\text{s}}}{\pi M_{\text{s}}^2 u})
/[\frac{u}{w}(1-2\ln(\frac{u}{w})+(\frac{\lambda_{\text{ex}}}{u})^2)]$ 
 depends both on the material and structural parameters ($K_{\text{s}}$ denotes the surface anisotropy). For $d \approx \infty$, as in our numerical calculations, $n =$ 0 means no nodal line in the upper or lower part of ADLW (see, Fig.~\ref{Fig:as_adw}; modes 1 and 2), $n=$ 1 a single nodal line (see, Fig~\ref{Fig:as_adw}; modes 3 and 4), etc.  The dispersion relation of SWs in the WG can be written in the form:\cite{Stnacil,Guslienko05}
$$
\omega =  \sqrt{\left( \omega_0 +\omega_{\text{ex}} \right) \left(  \omega_0 + \omega_{\text{ex}}+\omega_{\text{dip}} \right)},
$$
where $\omega$ is an angular frequency of SWs. $\omega_0=\gamma\mu_0 H_0$, $\omega_{\text{ex}}=\gamma\mu_0 M_{\text{s}}\frac{\lambda_{\text{ex}}^2}{w^2}(n^2\pi^2+k^{2}_x w^{2})$,  $\omega_{\text{dip}}=\gamma\mu_0 M_{\text{s}}\frac{1-\exp(k_x u)}{k_x u}$ denote the contributions from external, exchange and dipolar fields, respectively. 

The estimations of changes in SW dispersion relation resulting from the changes of $w$ or $H_{0}$ can be done by calculation of the full differential of the function $\omega=\omega(w,H_{\text{0}})$. It allows to derive the relation between small changes of $\Delta H_{0}$ and $\Delta w$, for which the compensation of intrinsic and extrinsic symmetry breaking can occur, i.e., when ${\text{d}}\omega(w,H_0)=$ 0: 
\begin{eqnarray}
\frac{\mu_0\Delta H_{0}}{\Delta w}&\approx& \frac{2\pi^2\mu_0  M_{\text{s}}\lambda_{\text{ex}}^2 (n+1)^2}{w^3}
\times 
\nonumber\\&  & 
f\!\!\left(\frac{K_{\text{s}}}{ \pi u \mu_0 M_{\text{s}}^2}-1,\frac{\lambda_{\text{ex}}}{w},\frac{u}{w}\right).\label{eq:ratio}
\end{eqnarray}
This ratio has units of T/m and shows how big asymmetric magnetic field needs to be added to compensate the shift of the anti-dots. 
The function:
$
f(s,l,r)=[s+\frac{1}{\pi}(2\frac{l^2}{r}-4r\ln(r))] [s+\frac{1}{\pi}(r+\frac{l^2}{r} - 2r\ln(r)) - s)]s^{-2}
$
depends on three dimensionless parameters:  $s$--the relative strength of the surface anisotropy, $l$--ratio between exchange length and the width of WG and $r$--the aspect ratio of the WG. 
The values of $f(s,l,r)$  with big absolute value of $K_{\text{s}}$, refer to the regime of strong pinning.  Note that Eq.~(\ref{eq:ratio}) does not depend on  $k_{x}$, which means that it should be fulfilled for any wave-vector. 

In our ADLW, we have two WGs separated by the anti-dots row. When we shift the row of the anti-dots $\Delta w$ along positive $y$ direction, the width of the upper WG is decreasing by $\Delta w$ and the width of the lower WG is increasing by the same amount.  It will result in an increase of the frequency of the modes concentrated in the upper WG and decrease of the frequency of the mode concentrated in the lower WG. 
To compensate for these changes in the dispersion relation  by a bias magnetic field we need to do the opposite. According to Eq.~(\ref{eq:ratio}) we need to apply different bias magnetic fields to upper and lower WG.
The dependence of the magnonic gap width under asymmetric bias magnetic field applied to the ADLW, $\Delta H_{0}$  (i.e., in the upper part of the ADLW bias magnetic field is $H_{0}+ \Delta H_{0}$, while in the lower part of ADLW is $H_{0}- \Delta H_{0}$) calculated with PWM is shown in the Fig.~\ref{Fig:Field}(b). 

\begin{figure}[!ht]
\includegraphics[width=8.5cm]{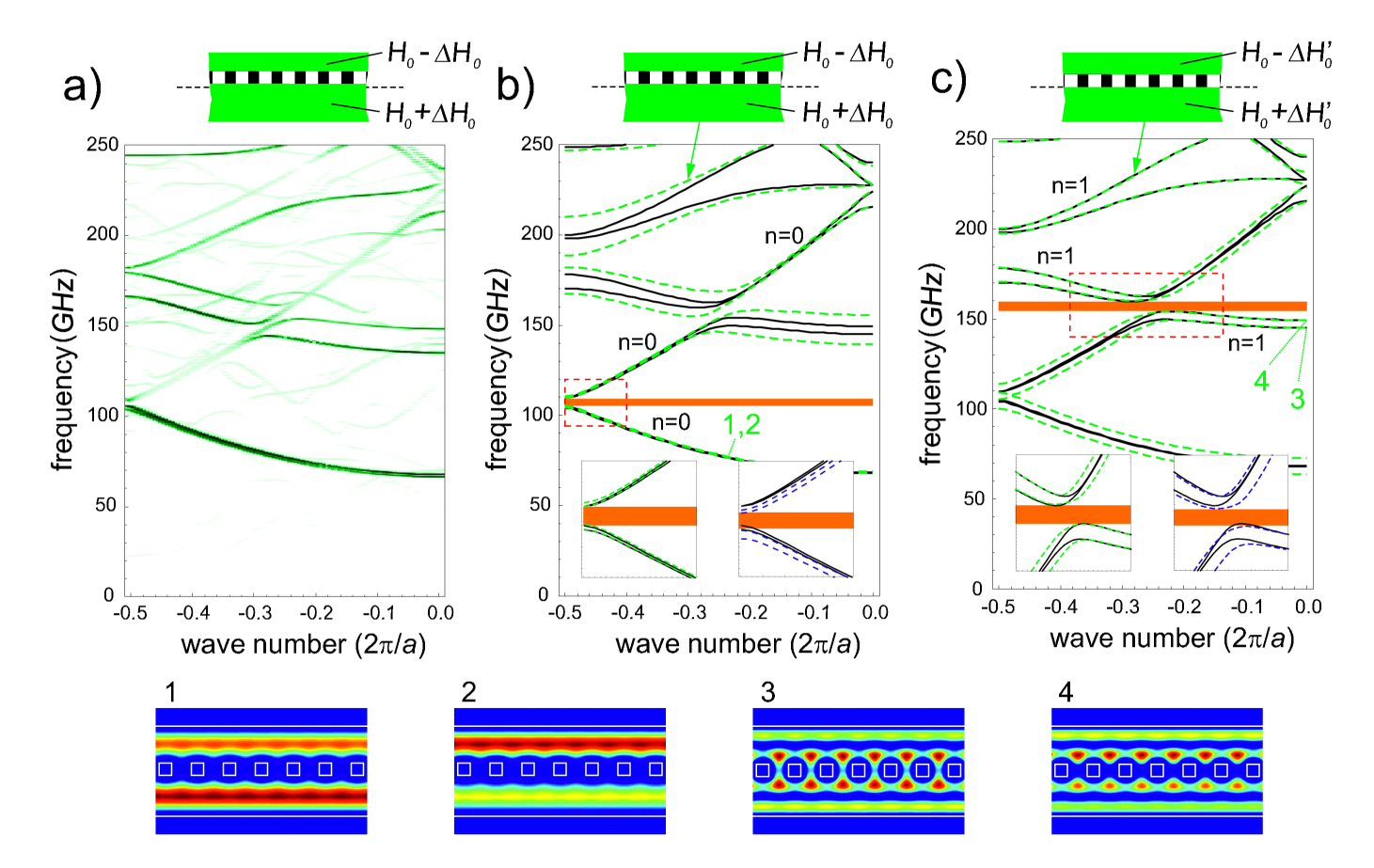}
\caption{(Color online) Magnonic band structure of the ADLW showing the compensation effect of the intrinsic asymmetry by the extrinsic field. In (a) and (b) (green dashed lines) the first gap (for the modes without a nodal line---$n=$ 0) is completely reopened. The calculations with OOMMF (a) and PWM (b) were performed for   $\Delta w = $ 1 nm and 0.9 nm, respectively, with $\mu_{0} \Delta H_{0} =$ 105 mT ($\mu_{0}H_{0} = 1$ T). The reopening of the second gap (opened in the anti-crossing of the mode $n$ = 0 with $n$ = 1) is presented in (c). Calculations in (c) were done with PWM for  $\mu_{0} \Delta H_{0} =$ 205 mT and $\Delta w = $ 0.5 nm. The left insets in (b) and (c) show enlarged results for the step-like field profile of the bias magnetic field, the right ones - the outcomes for linear change of the magnetic field profile (ramp-like profile) across the ADWL. At the bottom row, profiles of SW calculated with PWM are shown. Profiles for modes 1 and 2 are calculated for the structure from (b) and modes 3 and 4 for the structure from (c) at the BZ center. }
\label{Fig:Comp}
\end{figure}

After these estimations we perform the PWM calculations. The results are presented in the Fig.~\ref{Fig:Comp}(b) and (c) for $\Delta H_{0}$ to recover the first and the second magnonic gap in the asymmetric ADLW (i.e., when $\Delta w = 0.9$ nm and 0.5 nm), respectively (see Fig.\ref{Fig:Field}a). 
 It is interesting that it is possible to recover the first and the second magnonic gaps but with different values of the ratio $\frac{\mu_0\Delta H_{0}}{\Delta w}$. The analytical values of this ratio (calculated from the Eq.~(\ref{eq:ratio}) with $w$ = 18.5 nm, i.e., the distance between pinned layers used in MS) for the ideal pinning ($f(s,l,r)=1$) for the first gap (when $n=$ 0) and the second one ($n=$ 1) are $101$ mT/nm and $406$ mT/nm, respectively.  To validate our predictions we performed  MSs for $\mu_0\Delta H_{0} = $ 105 mT and  $\Delta w = $ 1 nm.
The MS results are shown in Fig.~\ref{Fig:Comp}(a) with the first frequency gap opened and with good agreement with PWM calculations shown in Fig.~\ref{Fig:Comp}(b). 
In order to open the second gap we have to cancel the shifts for the bands with a single nodal line ($n=$ 1) (this gap is formed due to the anti-crossing of the $n= 0$ and $n=1$ modes but at the $\Delta w$ = 0.5 nm the splitting of the $n=1$ dominate)  by applying the field for which  $\frac{\mu_0\Delta H_{0}}{\Delta w}$ is about 4 times bigger  than for the first gap (410 mT/nm). This confirms the applicability of Eq.~(\ref{eq:ratio}) with a square dependence on $n + 1$. The profiles of SWs (compare bottom panels of Fig.~\ref{Fig:as_adw} and Fig.~\ref{Fig:Comp}) further establish the restoration of amplitude distribution by extrinsic compensation.
Small differences in the extent of band gap recovery obtained from numerical calculations and the analytical model, show that the pinning in the middle of ADLW is not perfect.  


We have checked also that the compensation effect will still appear for continuous change of the bias magnetic field, which is also more justified in an experimental research. We used ramp-like profile of the magnetic field:\cite{Cho95} $H=H_0+2\Delta H_0(2y+\Delta w)$ where $y=0$ corresponds to the ADLW center. The values of $\Delta w$ and $\Delta H_0$ are the same as for step-like profile of magnetic field.
We have found, that this kind of the field acts similarly to the step-like profile of the field, when its value is normalize to have the same average value for corresponding sub-waveguides (the formula for ramp-like field fulfills this condition). The results of PWM calculations are shown in Fig.~\ref{Fig:Comp} in right insets, this was also confirmed in MSs. 

We have shown that a small symmetry braking in anti-dots waveguide can result in closing of magnonic gaps in the range of the spectra determined mainly by  exchange interactions. It can be crucial for practical realization of the SW waveguides because precise mirror symmetry will be difficult to achieve on such small scales. We have shown that the effect of a deviation from the ideal structure in magnonic band structure can be compensated by the bias magnetic field, i.e., in an extrinsic way. Using the predictions of an analytical model, we showed in numerical calculations that such a compensation is possible but the field should be asymmetric as well. It was presented  that two magnonic band gaps of different origins can be reopened in asymmetric waveguides by this way. Our predictions are based on the symmetry properties and a general analytical model, so it should be possible to extend this idea  to larger structures, i.e., closer to actual experimental samples where the dipolar interactions are important. The intrinsic and extrinsic symmetry breaking or its compensation can be also exploited to manipulate active and inactive modes, which couple to the external fields,\cite{Vlaminc10,Au12} in a similar way as was predicted for plasmonic metamaterials.\cite{Christ08}  Further study is necessary to investigate the compensation effects proposed here in other systems, e.g., like in electrons propagating in an periodically patterned graphene ribbons by the external electric field.\cite{Pedersen08,Oswald12}  

\begin{acknowledgments}
We acknowledge the financial support from the Department of Science and Technology, Government of India (Grant nos. INT/EC/CMS (24/233552), SR/NM/NS-09/2007), Department of Information Technology, Government of India (Grant no. 1(7)/2010/M\&C) and the European Community's FP7/2007-2013 (Grant Agreement nos. 233552 (DYNAMAG) and  228673  (MAGNONICS)).  D.K. would like to acknowledge financial support from CSIR - Senior Research Fellowship (File ID: 09/575/(0090)/2011 EMR-I). 
\end{acknowledgments}

\newpage

\end{document}